%% file: list-arxiv.tex
\documentclass[11pt,twocolumn]{article}

\usepackage[bitstream-charter,greekuppercase=upright,cal=cmcal]{mathdesign}
\usepackage[T1]{fontenc}

\usepackage[sumlimits]{amsmath}
\usepackage{amsthm,amsopn,mathabx}
\usepackage[normalem]{ulem}
\usepackage{multirow}
\usepackage{units}
\usepackage{graphicx}
\usepackage{enumitem} 
\usepackage[english]{babel}

\usepackage[margin=2cm]{geometry}
\usepackage{url}
\usepackage{verbatim}
\usepackage{xcolor}
\definecolor{dullmagenta}{rgb}{0.5,0,0.4}
\definecolor{dullblue}{rgb}{0.0,0,0.86}

\allowdisplaybreaks

\newtheorem{theorem}{Theorem}[section]

\newtheorem{lemma}[theorem]{Lemma}

\usepackage[noindentafter]{titlesec}

\titleformat*{\section}{\large \bfseries}
\titleformat*{\subsection}{\normalsize\bfseries}

\titlespacing\section{0pt}{12pt plus 4pt minus 2pt}{2pt plus 2pt minus 2pt}

\usepackage[backend=biber,sorting=none,style=phys-jmr,biblabel=brackets,backref=true,bibencoding=utf8,maxnames=20,eprint=true,pageranges=false]{biblatex}
\makeatletter
\pretocmd{\blx@head@bibintoc}{\phantomsection}{}{\ddt}
\makeatother
\DefineBibliographyStrings{english}{%
  backrefpage = {page},
  backrefpages = {pages},
}
\usepackage{csquotes}

\usepackage[ruled]{algorithm2e}
\usepackage{hyperref}
\hypersetup{colorlinks=true,citecolor=dullblue,linkcolor=dullblue,filecolor=dullblue,urlcolor=dullblue,breaklinks=true,bookmarksnumbered=true, bookmarksopen=true,bookmarksopenlevel=1}

\usepackage{tikz}
\usetikzlibrary{positioning}
\definecolor{blueviolet}{RGB}{138,43,226}
\definecolor{darkgray176}{RGB}{176,176,176}
\definecolor{lightgray211}{RGB}{211,211,211}
\definecolor{darkorange25512714}{RGB}{255,127,14}
\definecolor{dodgerblue}{RGB}{30,144,255}
\definecolor{gold}{RGB}{255,215,0}
\definecolor{red}{RGB}{255,0,0}
\definecolor{orange}{RGB}{255,93,0}
\definecolor{limegreen}{RGB}{50,205,50}
\definecolor{steelblue31119180}{RGB}{31,119,180}
\definecolor{teal}{RGB}{0,128,128}
\definecolor{olivegreen}{RGB}{55,126,34}
\definecolor{neonpurple}{RGB}{176,38,255}
\definecolor{coral}{RGB}{254,125,106}
\definecolor{lemonade}{RGB}{252,186,203}
\definecolor{lemon}{RGB}{255,247,0}
\definecolor{amber}{RGB}{255,191,0}
\definecolor{palegreen}{RGB}{180,238,180}
\definecolor{lightteal}{RGB}{64,224,208}
\definecolor{lavender}{RGB}{227,159,246}

\usepackage{pgfplots}
\pgfplotsset{compat=newest}
\usepgfplotslibrary{groupplots}
\usepgfplotslibrary{dateplot}
\newcounter{plot}[figure]

\usepackage{cleveref}
\crefname{plot}{plot}{plots}
\Crefname{plot}{Plot}{Plots}

\usetikzlibrary{quantikz2}
\usepackage{floatrow, float}
\usepackage[label font=bf,labelformat=simple]{subfig}

\usepackage[font=small,skip=0pt,margin=10pt]{caption}

\usepackage{braket}

\usepackage[title]{appendix}

\newcommand\blfootnote[1]{%
  \begingroup
  \renewcommand\thefootnote{}\footnote{#1}%
  \addtocounter{footnote}{-1}%
  \endgroup
}

\input{macrosetup}

\usepackage{setspace}

\usepackage{titling}

\pretitle{\begin{center} \large \bfseries}
\posttitle{\par\end{center}}
\predate{}
\postdate{}
\renewcommand{\abstractname}{\vspace{-\baselineskip}}
\preauthor{\begin{center} \normalsize}
\postauthor{\end{center}}

\begin{document}

\title{Graph Neural Networks for Enhanced Decoding of Quantum LDPC Codes}
\author{Anqi Gong$^1$, Sebastian Cammerer$^2$, and Joseph M.~Renes$^1$\\
{\itshape $^1$Institute for Theoretical Physics, ETH Z\"urich, 8093 Z\"urich, Switzerland\\
$^2$NVIDIA
}
}

\date{}
\renewcommand{\abstractname}{\vspace{-1.25\baselineskip}}

\twocolumn[{%
  \begin{@twocolumnfalse}
    \maketitle

\begin{abstract}
In this work, we propose a fully differentiable iterative decoder for quantum low-density parity-check (LDPC) codes. The proposed algorithm is composed of \emph{classical} belief propagation (BP) decoding stages and intermediate graph neural network (GNN) layers. Both component decoders are defined over the same sparse decoding graph enabling a seamless integration and scalability to large codes.
The core idea is to use the GNN component between consecutive BP runs, so that the knowledge from the previous BP run, if stuck in a local minima caused by trapping sets or short cycles in the decoding graph, can be leveraged to better initialize the next BP run.
By doing so, the proposed decoder can learn to compensate for sub-optimal BP decoding graphs that result from the design constraints of quantum LDPC codes. 
Since the entire decoder remains differentiable, gradient descent-based training is possible. We compare the error rate performance of the proposed decoder against various post-processing methods such as random perturbation, enhanced feedback, augmentation, and ordered-statistics decoding (OSD) and show that a carefully designed training process lowers the error-floor significantly. As a result, our proposed decoder outperforms the former three methods using significantly fewer post-processing attempts. The source code of our experiments is available online.
\end{abstract}
\vspace{5mm}

  \end{@twocolumnfalse}
}]

\section{\label{sec:intro}Introduction}
Quantum low-density parity-check (QLDPC) codes are among the most promising types of error correction codes for fault-tolerant quantum computing. In recent years there have been numerous efforts to design QLDPC codes of high rate and high distance \cite{balanced_product,almost_linear_distance}; even linear distance has been investigated in \cite{asmptotically_good_panteleev,quantum_tanner_codes}. Recent work from Bravyi et al.~\cite{high_threshold} shows how certain codes can be mapped to a bilayer hardware architecture, spurring further interest in the community and increasing the practical importance of low-complexity decoders for such codes.\blfootnote{\url{https://github.com/gongaa/Feedback-GNN}} 

In this work, we focus on decoding medium block-length and high-rate CSS QLDPC codes introduced in \cite{panteleev_degenerate}. 
These codes are of particular interest due to their strong performance under maximum likelihood decoding and its approximations, such as OSD; however, their performance is known to be suboptimal under plain BP decoding. This can be intuitively explained by the fact that the corresponding Tanner graph contains unavoidable 4-cycles \cite{15years} due to the stabilizer commutation requirement. 
On the other hand, since errors differing by a stabilizer have the same syndrome, symmetric stabilizer trapping sets \cite{perturbation,trapping_set_QLDPC,stabilizer_inactivation}, where BP cannot decide between certain such errors (e.g., two equal-weight ones), also degrades BP performance.
The authors in \cite{perturbation} proposed to alleviate this issue by adding a small perturbation to the channel prior probabilities given as input to the BP decoder. An improved version is shown in \cite{enhanced_feedback}, where the perturbation depends on the unsatisfied checks. 

\begin{figure}[tb]
\centering 
\input{block_diagram}
\vspace*{-3mm}
\caption{\label{fig:block_diagram}
Block diagram of the proposed decoder architecture consisting of trainable GNN layers (orange) and \emph{classical} BP iterations (yellow). The same GNN is sandwiched between block BP runs of iteration $(64,16,16,16)$.}
\end{figure}
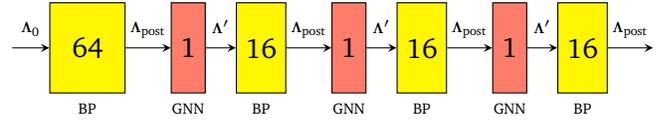

The goal of our work is to learn those perturbations using a GNN. The GNN follows the concept of \cite{cammerer2022graph} and acts as an intermediate layer between independent BP runs, as shown in \Cref{fig:block_diagram}. It takes the output log-likelihood ratios (LLRs) $\mathbf{\Lambda}_{\text{post}}$ estimated by the previous BP decoding stage, then calculates the reliabilities of checks, and uses both information to provide a refined channel prior $\mathbf{\Lambda'}$ as initialization of the next BP run. 

We call this intermediate layer a \emph{feedback} GNN. The GNN operates on the same decoding graph as the BP decoder, facilitating a seamless integration with the BP decoding stages. 
Our trained feedback GNN requires significantly fewer attempts than the aforementioned two methods \cite{perturbation,enhanced_feedback} to achieve a lower logical error rate, as shown in Fig. (\ref{fig:1270_and_882}\textbf{a}).

Machine learning for channel decoding is doomed by the curse of dimensionality meaning that the training complexity scales exponentially with the codeword length \cite{gruber2017deep}. However, we would like to emphasize that one advantage of the proposed feedback GNN structure is that the graph is explicitly provided and the main decoding task still relies on BP. Thereby, its training complexity is only mildly affected by the codeword length \cite{cammerer2022graph}.

The remainder of the paper is organized as follows. Section \ref{sec:LDPC} contains some background information on the CSS-type LDPC codes, and in Section \ref{sec:existing_postprocessing}, we review some existing post-processing methods. Section \ref{sec:BP} reviews the quaternary BP decoding method. With this basis we describe the feedback GNN model architecture and the training method in Section \ref{sec:feedback_GNN}. Section \ref{sec:simulation} presents the numerical simulation results. We conclude in Section \ref{sec:conclusion}.

\section{Quantum Low-Density Parity-Check Codes}
\label{sec:LDPC}

In the following we restrict our attention to the i.i.d.\ depolarizing channel, in which each qubit independently and identically experiences a random $X$, $Z$, or $Y=iXZ$ type of error, each with probability $p/3$ for some $p\in [0,1]$. 
\subsection{CSS Quantum Codes}
Quantum Calderbank-Shor-Steane (CSS) \cite{CSS-CS, CSS-Steane} codes are constructed from two \emph{classical} codes $C_X=\ker H_X$ and $C_Z=\ker H_Z$ defined by parity-check matrices $H_X$ and $H_Z$ under the requirement $H_X H_Z^T=\mathbf{0}$, i.e., $C_Z^\perp\subseteq C_X$ or equivalently $C_X^\perp \subseteq C_Z$. 
Rows of $H_X$ and $H_Z$ form the $X$- and $Z$-type stabilizer generators, while the $X$-type and $Z$-type logical operators are $C_Z/ C_X^\perp$ and $C_X/ C_Z^\perp$, respectively. All parity-check operations are binary operations in GF(2).

If $C_X$ is a blocklength $N$ code, encoding $K_X$ info bits (an $[N,K_X]$ code), and $C_Z$ is an $[N,K_Z]$ code, then the resulting quantum code is an $\llbracket N,K_X+K_Z-N\rrbracket$ code. The quantum minimum distance of this code is $d=\min\{d_X,d_Z\}$, where $d_X=\min\{\wt(\bx), \bx\in \ker H_Z/ \im H_X^T\}$, $d_Z=\min\{\wt(\bx), \bx\in \ker H_X/ \im H_Z^T\}$. $\wt(\bx)$ is the Hamming weight of the binary vector $\bx$.

An error is specified by $N$-bit vectors $\be_X$ and $\be_Z$ which denote the $X$ and $Z$ components of the error, respectively. 
A decoder takes the observed syndromes $\bs_X=H_Z\be_X$ and $\bs_Z=H_X\be_Z$ and constructs correction operations $\hbe_X$ and $\hbe_Z$. 
Importantly, these do not need to precisely match the error for decoding to succeed.  
Instead, $\hbe_X$ and $\be_X$ may differ by an $X$-stabilizer, and $\hbe_Z$ and $\be_Z$ may differ by a $Z$-stabilizer, as these correction operations will return the system to the codespace without incurring any logical error. 

The requirement for successful correction can be compactly written as $H_X^{\perp} (\hbe_X+\be_X)=\mathbf{0}$ and $H_Z^{\perp} (\hbe_Z+\be_Z)=\mathbf{0}$. 
This is more stringent than only requiring $\hbe$ to have the correct syndrome, i.e., $H_Z\hbe_X=H_Z\be_X$ and $H_X\hbe_Z=H_X\be_Z$, since the row span of $H_X$ is contained in the row span of $H_Z^{\perp}$ for a CSS code. 

In the context of the depolarizing model, the $X$ and $Z$ components of the CSS codes can be separately decoded using classical (binary) syndrome BP decoders. 
To address the correlation between $X$ and $Z$ errors, a \emph{quaternary} BP decoder \cite{15years} was introduced to directly operate on the Tanner graph constructed using both the $X$ and $Z$ checks. However, due to the CSS code constraints, this Tanner graph inevitably contains cycles of length 4 \cite{15years}. This renders the decoding of QLDPC codes into a challenging, but from research perspective yet attractive, task. 

\subsection{Existing Post-Processing Methods}\label{sec:existing_postprocessing}

One difficulty of decoding QLDPC codes using message passing methods is that the correction operator $\hbe=(\hbe_X|\hbe_Z)$ of the decoder is not assured to result in a zero syndrome $\bs=(\bs_X|\bs_Z)=\mathbf{0}$. 
We call such cases ``flagged'' events. 
To overcome such flagged events, several post-processing methods have been proposed.
One approach is to take the output of the BP decoder and use it to adjust the inputs to another round of BP decoding, in the hope that that round will lead to an unflagged error. 
The following is a summary of such methods, which can be applied repeatedly until the parity-checks are satisfied or the maximal number of attempts $N_a$ is reached:
\begin{itemize}[leftmargin=*]
    \item Random perturbation \cite{perturbation} randomly chooses unsatisfied (``frustrated'') checks and randomly perturbs the input LLRs for all the qubits involved in those checks. 
    \item Enhanced feedback \cite{enhanced_feedback} randomly chooses one frustrated parity-check and one of its associated qubits. The channel prior of this qubit is adjusted such that an anticommuting/commuting error is more likely if the ground truth syndrome of the parity-check is 1/0. 
    \item Section III.A of \cite{augmentation} contains an overview of the first two methods. In addition, \cite{augmentation} proposes a matrix augmentation method, which randomly chooses and repeats some of the rows from the parity-check matrix, and uses this augmented matrix to initialize the message passing decoding in the next round.
    \item Stabilizer inactivation (SI) \cite{stabilizer_inactivation} ranks the parity-checks according to their reliability, as measured by the sum of the absolute values of the posterior LLRs of the associated qubits. Then the qubits associated to the most unreliable checks are ``inactivated'' by taking them out in the next round of message passing decoding. Finally, the error on the inactivated qubits is determined by solving a small system of linear equations. 
    
\end{itemize}

A related method is \emph{zeroth}-order OSD \cite{panteleev_degenerate}, which is applied only once after BP decoding. By its nature, OSD always gives an estimate that satisfies the syndrome equations, thus eliminating flag errors. This involves solving a rank$(H)$ system of linear equations. Higher-order OSD was also introduced in \cite{panteleev_degenerate}, but is beyond of the scope of this work. 

\section{Quaternary Belief Propagation Decoder}
\label{sec:BP}
The quaternary BP (BP4) algorithm utilizes the correlation between $X$ and $Z$ noise by directly working on the quaternary alphabet $\{I,X,Y,Z\}$.
Here we discuss the details of the algorithm, as they will be useful later. 
As with usual BP decoding algorithms, BP4 utilizes the Tanner graph of the code, and proceeds by sending messages back and forth between VNs to CNs. The goal is to estimate, for each qubit, the marginal probability of error on that qubit, given the observed syndrome. 

Following \cite{refinedBP,log_domain}, the input to the BP4 algorithm is the syndrome pair $\bs_X=H_Z\be_X$ and $\bs_Z=H_X\be_Z$. 
Initialize variable nodes to $\mathbf{\Lambda}\in\R^{N\times 3}$, where the $i^{th}$ row $\mathbf{\Lambda_i}=(\LX_i,\LY_i,\LZ_i)$ is the sequence of LLRs associated with $X$, $Y$, and $Z$, respectively. 
Specifically, $\LX_i=\log{\frac{p^I_i}{p^X_i}}$, where $p_i^I$ and $p_i^X$ are the probabilities of no error or an $X$ error happening on VN $i$, respectively. 
In the case of i.i.d.\ depolarizing noise $p_i^X=p_i^Y=p_i^Z=p/3$ for all VN $i$, meaning the initial VN LLRs are 
\begin{equation}\label{eq:initialization}
    \LX_i=\LY_i=\LZ_i=\log{\frac{1-p}{p/3}}\quad \forall i\,.
\end{equation}

The message $\lambda_{i\to j}$ sent from VN $v_i$ to CN $c_j$ is a scalar. For example, if $c_j$ involves an X-type check on $v_i$. Then this scalar LLR message is 
\begin{equation}\label{eq:mapping}
\lambda_{i\to j}=\log\frac{p_i^I+p_i^X}{p_i^Y+p_i^Z}=\log\frac{1+e^{-\LX_i}}{e^{-\LY_i}+e^{-\LZ_i}}\,
\end{equation}
since either $Y$ or $Z$ error contributes $1$ to (anti-commutes with) the check. 
When $c_j$ involves a $Y$- or a $Z$-type check the messages are defined similarly, as the LLR of an anticommuting error. 
This mapping (and thus the BP4 algorithm) can also be applied to non-CSS codes, however, we restrict ourselves to CSS QLDPC codes in this work. As a consequence, every check node involves either pure $X$ checks or pure $Z$ checks on VNs.

\subsection{Check Node Update}
The message $\Delta_{j\to i}$ from CN $c_j$ back to VN $v_i$ is the LLR of whether $v_i$ commutes with $c_j$, conditioned on the check node observations. 

\begin{lemma}\label{lem:Gallager}{\textbf{Gallager \cite{Gallager_ldpc}.}}
Consider a sequence of $m$ independent binary digits $(a_1,...,a_m)$ in which $\Pr[a_k=1]=p_k$. Then the probability $q$ that $a_1\oplus\dots \oplus a_m=1$ (i.e., an odd number of ones occurring) is 
\begin{equation}\label{eq:Gallager}
q=\frac{1}{2}-\frac{1}{2}\prod_{k=1}^m (1-2p_k)\,
\end{equation}
which can be rewritten in the log-domain as
\begin{equation}\label{eq:Gallager_boxplus}
\log{\frac{1-q}{q}}=2 \tanh^{-1}{\left[\prod_{i=1}^m \tanh\left( \frac{1}{2} \log{\frac{1-p_i}{p_i}}\right)\right]}\,.
\end{equation}
\end{lemma}
The right-hand side is usually abbreviated as the boxplus operation $\boxplus_{i=1}^m \Lambda_i$ for LLRs $\Lambda_i=\log{\frac{1-p_i}{p_i}}$. Later, the loss function and the feature required by the GNN both rely on this simple lemma.

The message $\Delta_{j\to i}$  involves extrinsic information only, and is given by 
\begin{equation}\label{eq:CN_update}
\Delta_{j\to i}=(-1)^{s_j}\cdot \boxplusop_{i'\in \mathcal{N}(j)\backslash\{i\}} \lambda_{i'\to j}
\end{equation}
where $s_j\in \{0,1\}$ is the syndrome of the check, $\mathcal{N}(j)$ is the set of neighboring VNs of $c_j$ in the Tanner graph. 
\subsection{Variable Node Update}
The VN update processes the messages coming from the CNs in the direct neighborhood.
Assume $c_j$ involves an X-type check on $v_i$, the LLR message $\Delta_{j\to i}$ cannot distinguish between $I$ and $X$ noise as they both commute with the check, similarly, it cannot distinguish $Y$ and $Z$ noise. Therefore, it is assumed that those posterior probabilities $p^{j\to i}_I=p^{j\to i}_X$ and $p^{j\to i}_Y=p^{j\to i}_Z$, and hence $\Delta_{i\leftarrow j}=\log\frac{p^{j\to i}_I+p^{j\to i}_X}{p^{j\to i}_Y+p^{j\to i}_Z}=\log\frac{p^{j\to i}_I}{p^{j\to i}_Z}$. Similarly, if $c_j$ involves a Z-type check on $v_i$, then $\Delta_{i\leftarrow j}=\log\frac{p^{j\to i}_I}{p^{j\to i}_X}$.

The update of the the LLR vector $(\Gamma_{i\to j}^X, \Gamma_{i\to j}^Y, \Gamma_{i\to j}^Z)$ for $v_i$ also done via the extrinsic information rule given as
\begin{equation}\label{eq:X_VN_update}
\Gamma_{i\to j}^X=\Lambda_i^X+\sum_{j'\in \mathcal{M}_Z(i)\backslash \{j\}} \Delta_{i\leftarrow j'}
\end{equation}
where $\mathcal{M}_Z(i)$ are all the $Z$-type check nodes (CSS code) involving $v_i$. Note that checks $c_j$ from $\mathcal{M}_X(i)$ make no contribution as $\log\frac{p_I^{j\to i}}{p_X^{j\to i}}=0$.
Similarly, the update rule for $\Gamma_{i\to j}^Y$ is 
\begin{equation}\label{eq:Y_VN_update}
\Gamma_{i\to j}^Y=\Lambda_i^Y+\sum_{j'\in \mathcal{M}_Z(i)\bigcup \mathcal{M}_X(i)\backslash \{j\}} \Delta_{i\leftarrow j'}\,.
\end{equation}
Using \eqref{eq:mapping}, one can combine messages to send to $X$ and $Z$-type CNs and continue the BP iterations.

\subsection{Hard Decision}
Before a final hard decision, the last decoding iterations calculate the channel posterior 
$\mathbf{\Lambda}_{\text{post}}=(\mathbf{\Gamma}^X,\mathbf{\Gamma}^Y,\mathbf{\Gamma}^Z)$
whose individual components read
\begin{equation}\label{eq:channel_posterior}
\begin{split}
\Gamma_i^{X/Z} &= \Lambda_i^{X/Z}+\sum_{j'\in \mathcal{M}_{Z/X}(i)} \Delta_{i\leftarrow j'}\\
\Gamma_i^Y &= \Lambda_i^Y+\sum_{j'\in \mathcal{M}_Z(i)\bigcup \mathcal{M}_X(i)} \Delta_{i\leftarrow j'}
\end{split}
\end{equation}
and $\Gamma_i^I=0$ and then for each VN choose among $\{I,X,Y,Z\}$ that leads to the smallest LLR.

\section{Feedback GNN}\label{sec:feedback_GNN}

\begin{figure}[ht]
\centering 
 \input{fb_gnn}
 \vspace*{-3mm}
 \caption{\label{fig:tanner_graph}
Unrolled feedback GNN operating on the Tanner graph, showing the inside of the orange boxes in Fig. (\ref{fig:block_diagram}). The VN feature is initialized using $\mathbf{\Lambda}_{\text{post}}$ from the previous BP run and the CN feature is calculated using Eq (\ref{eq:combination_explicit},\ref{eq:CN_feature}). Each edge message is calculated using the features of its two endpoints. After that, each variable node aggregates the incoming X (red) and Z-type (blue) messages and then uses them together with its own feature to obtain a modified channel prior $\mathbf{\Lambda'}$ for the next BP run. Solid objects indicate trainable operations.
 }
\end{figure}
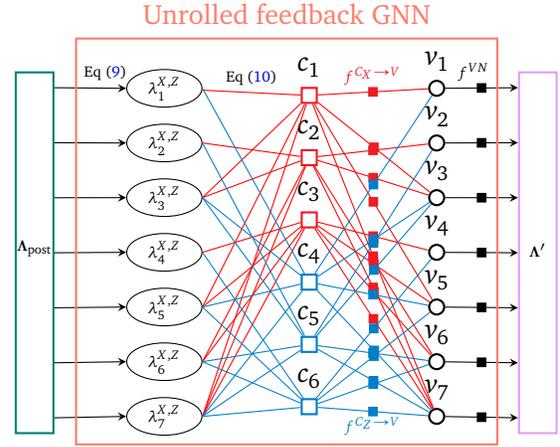

The proposed message passing GNN decoder \cite{cammerer2022graph} utilizes the same Tanner graph as the BP decoder, but differs in the representation of the messages and uses trainable CN and VN update functions which are parameterized, for instance by using simple MLP layers. 
Each variable node or check node has an assigned feature vector. The message on an edge is computed using the concatenated features of its two end nodes. Afterward, the state of a node is updated using its previous state and by the aggregated incoming messages from its neighborhood. The aggregation operation used in this work is the mean operation.
To keep the decoding complexity low, our feedback GNN eliminates the ``iterative'' aspect of the decoder, opting for just one VN update iteration. However, the basic concept still follows \cite{cammerer2022graph}.

\subsection{Architecture}\label{sec:fb_GNN_architecture}
After the BP run, the posterior channel LLR is obtained using \eqref{eq:channel_posterior} and it becomes the variable node feature $\mathbf{h}_{v_i}=\mathbf{\Lambda}_{\text{post},i}=(\Gamma_i^X, \Gamma_i^Y, \Gamma_i^Z)$. Using \eqref{eq:mapping}, the two LLRs $\lambda_i^X, \lambda_i^Z$ of whether $v_i$ commutes with X or Z check are
\begin{equation}\label{eq:combination_explicit}
   \lambda_i^X=\log\frac{1+e^{-\Gamma_i^X}}{e^{-\Gamma_i^Y}+e^{-\Gamma_i^Z}},\quad \lambda_i^Z=\log\frac{1+e^{-\Gamma_i^Z}}{e^{-\Gamma_i^X}+e^{-\Gamma_i^Y}}\,. 
\end{equation}

Using Lemma~\ref{lem:Gallager}, the LLR of whether X-type check node $c_j$ is satisfied is $\boxplusop_{i'\in N(j)}\lambda^Z_{i'}$.
The feature of this check node is 
\begin{equation}\label{eq:CN_feature}
   h_{c_j}=(-1)^{s_j}\times \boxplusop_{i'\in N(j)}\lambda^Z_{i'}
\end{equation}
which is a scalar, and the more negative it is, the more likely this check tends to be not satisfied.

Next, for each CN to VN edge, concatenate the features of its two endpoints and calculate the message
\begin{equation}\label{eq:GNN_edge}
    \mathbf{m}_{c_j\to v_i}=f^{C_{X/Z}\to V}\left(\left[h_{c_j}||\mathbf{h}_{v_i}\right], \boldsymbol{\theta}_{C_{X/Z}\to V}\right).
\end{equation}
X-type CNs share weights $\boldsymbol{\theta}_{C_{X}\to V}$ for the function $f^{C_X\to V}$. Z-type CNs share weights $\boldsymbol{\theta}_{C_{Z}\to V}$ for $f^{C_Z\to V}$. In our implementation $f^{C_X\to V}$ and $f^{C_Z\to V}$ are both MLPs with one hidden unit of 40 neurons and \emph{tanh} activation. Both MLPs project the messages to dimension 20 as output.

For each $v_i$, calculate the average of all the incoming messages from X-type and Z-type checks respectively.
\begin{equation}\label{eq:reduce_msg}
\begin{split}
\mathbf{m}^X_{v_i} &= \frac{1}{|\mathcal{M}_X(i)|} \sum_{j'\in \mathcal{M}_X(i)} \mathbf{m}_{c_j'\to v_i}\\
\mathbf{m}^Z_{v_i} &= \frac{1}{|\mathcal{M}_Z(i)|} \sum_{j'\in \mathcal{M}_Z(i)} \mathbf{m}_{c_j'\to v_i}
\end{split}
\end{equation}
All VNs share weight for $f^{VN}$, which is another MLP (hidden unit 40) that projects the high-dimensional messages back to three dimensions. 
\begin{equation}\label{eq:GNN_VN}
    \mathbf{\Lambda'}_{i}=f^{VN}\left(\left[\mathbf{h}_{v_i} ||\mathbf{m}^X_{v_i} ||\mathbf{m}^Z_{v_i} \right], \boldsymbol{\theta}_{VN}\right)\,.
\end{equation}
The resulting output $\mathbf{\Lambda}'\in \R^{N\times 3}$ is then used to initialize the next BP run.

This feedback GNN essentially only does one VN update, in contrast to the usual message-passing GNN \cite{cammerer2022graph} that does multiple iterations of CN and then VN updates. Moreover, the proposed feedback GNN is lightweight, containing just 3923 trainable parameters in total, irrespective of the code size.

\subsection{Boxplus Loss}\label{sec:boxplus_loss}
In order to train this GNN, we feed its output $\mathbf{\Lambda}'$ to initialize another BP4 decoding stage, and let it run for 16 iterations. We record $(\Gamma_i^X,\Gamma_i^Y,\Gamma_i^Z)$ as in Eq \eqref{eq:channel_posterior} at the end of iteration 8 to 16 and calculate $\lambda_i^X, \lambda_i^Z$ for all these iterations as in Eq \eqref{eq:combination_explicit}, and for all X/Z-type checks calculate $\boxplusop_{i'\in \mathcal{N}(j)} \lambda_{i'}^{Z/X}$, respectively. These are the logits used to calculate a binary cross entropy loss with the ground truth syndrome $s_j$. This loss is minimized when $H_X \hbe_Z=H_X \be_Z$ and $H_Z \hbe_X=H_Z \be_X$.

The proposed \emph{boxplus} loss is different from the sine loss introduced in \cite{NBP}. It is also possible to extend this boxplus loss to take degeneracy into account.

For CSS codes, no logical error happens if and only if $H_X^{\perp} \hbe_X=H_X^{\perp} \be_X$ and $H_Z^{\perp} \hbe_Z=H_Z^{\perp} \be_Z$. And the CSS requirement implies $H_Z\subset H_X^{\perp}$  and $H_X\subset H_Z^{\perp}$. Therefore, we can simply enlarge the X and Z parity-check matrices into $H_Z^{\perp}$ and $H_X^{\perp}$ when doing the multi-loss calculation.
When training the feedback GNN, we do not take degeneracy into account. 

\subsection{Training methods}\label{sec:training_methods}
\begin{itemize}[leftmargin=*]
    \item Step 1: Generate a dataset using BP4 (64 iterations, flooding), and store the error patterns $(\be_x,\be_z)$ for which BP4 fails to decode. Only $\be_x \lor \be_z$  (bit-wise OR) having Hamming weight up to $0.05\times n$ are considered. We gather roughly a million such errors.
    \item Step 2: Use this dataset to train a coarse GNN which is then used to generate \emph{hard-to-decode} examples in the next step. The GNN is embedded between two BP runs of 16 iterations. The output of iterations 8 to 16 of the latter run is used to calculate a boxplus multi-loss, as described in the last subsection. Training takes roughly ten minutes on a single NVIDIA RTX4090.
    \item Step 3: Use the trained coarse GNN embedded into two 64-iteration BP components to generate a more sophisticated dataset. Note that we need fewer samples from this dataset when compared to step 1 (empirically 1/50 of the original size).
    \item Step 4: We mix the \emph{hard-to-decode} dataset with the easy one and balance the probability of occurrence (such that easy/hard to decode samples are equally likely). We then finetune the GNN sandwiched between (64,16) BP blocks for the new dataset. Each training takes less than half an hour.

\end{itemize}
Note that we train ten such models in parallel and select the best one. 

We implement the BP decoder together with a normalization factor $\kappa$. Specifically, the messages from CN to VN in Eq (\ref{eq:X_VN_update},\ref{eq:Y_VN_update}) are multiplied by this scaling factor. During training, the factor $\kappa=1.0$ is used in every BP run (i.e.\ no scaling). However, we empirically observed that during evaluation, at high physical error rates, it is beneficial to set $\kappa$ for the first run of BP to $0.8$. It is also interesting to observe that the feedback GNN improves the logical error rate even at high physical error rates $p$ though the model was not trained there. However, we do not show the curves at low physical error rates for $\kappa=0.8$, as they have higher error floors when compared to $\kappa=1.0$.

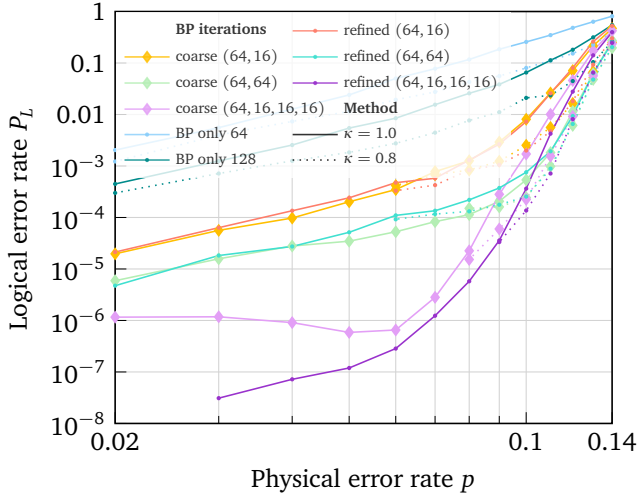
\begin{figure}[htb]
\input{n1270_progression}
\caption{\label{fig:n1270_k28_progression} Logical error rate of $\llbracket 1270,28,\leq 46 \rrbracket$ codes using feedback GNNs on depolarizing channel. Comparison of the performance of the coarse and the refined GNN trained on easy and mixed samples respectively.}
\end{figure}

The GNN is trained by being embedded between two BP blocks, but during evaluation, it is reused for possibly more than one attempt. Let (64,16,16,16) denote the decoder consisting of 64 BP iterations followed by $3 \times 16$ BP iterations each with an intermediate GNN layer sharing the same weight.
In \Cref{fig:n1270_k28_progression}, one can see three attempts (64,16,16,16) are better than one attempt (64,64), though the latter involves more total BP runs but less GNN layers. 

We choose (64,16,16,16) to be the final decoder. Hereby, steps 3 and 4 of the training improve the performance at low $p$. As can be seen in \Cref{fig:n1270_k28_progression}, using the coarse GNN has an error floor of a logical error rate around $10^{-6}$ while the finetuned version shows a significantly lower error floor.

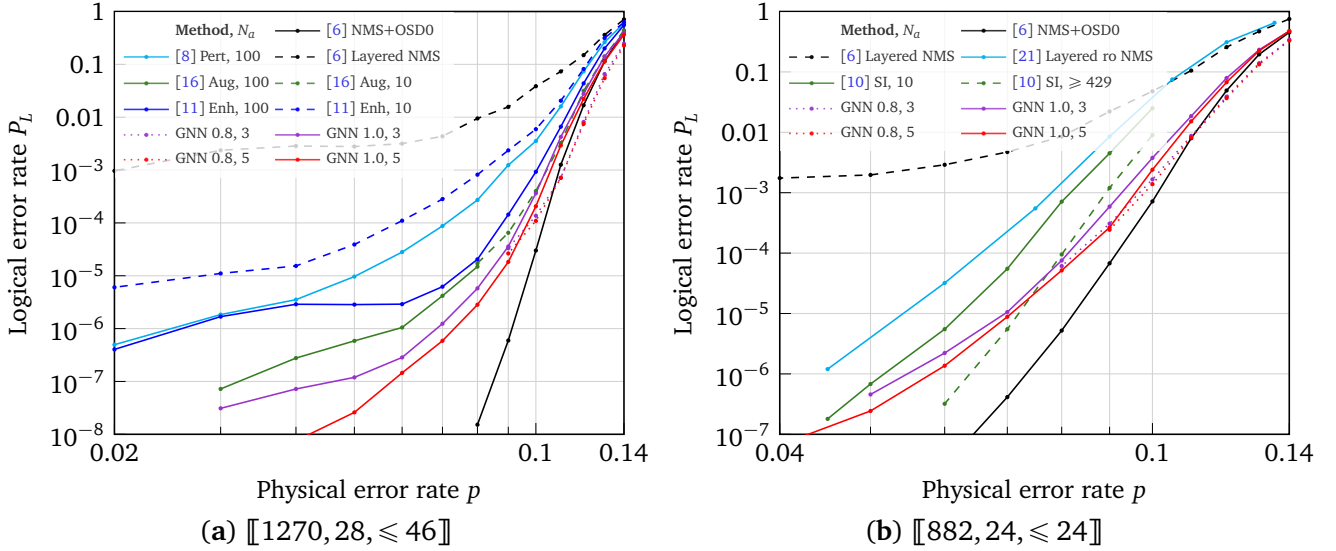
\begin{figure*}[htb]
\centering
\input{n1270_k28}        
\vspace{2pt}
\input{n882_k24}        
\caption{Logical error rate of the $\llbracket 1270,28, \leq 46\rrbracket$ and the $\llbracket 882,24,\leq 24 \rrbracket$ codes using various post-processing methods on depolarizing channel. $N_a$ is the maximum number of attempts. For our feedback GNNs, only the first block run of BP4 needs 64 iterations, while 16 iterations are enough for the post-processing block BP4 run. For example, three attempts will involve $64+16\times 3=112$ iterations of flooding BP in total. The factor $\kappa=1.0$ or $0.8$ was used for the first block run of BP, all later runs used $1.0$. (\textbf{a}) All the curves except the four feedback GNN ones are taken from Fig. (3) of \cite{panteleev_degenerate}, where a 32-iteration layered normalized min-sum (NMS) decoder with factor $0.625$ was used as the MP decoder in every attempt. Our feedback GNN, which aims to capture the core ideas of random perturbation \cite{perturbation} and enhanced feedback \cite{enhanced_feedback}, at three attempts indeed outperforms these two methods repeated 100 times. (\textbf{b}) The random order layered binary perturbed NMS \cite{layered} method was implemented on Z-noise, converted here to depolarizing noise using the $2/3$ rule. Their total number of steps needed is $64\times 3.5=224$ for Z noise. While ours is at most $64+16\times 5=144$ for depolarizing noise. The stabilizer inactivation \cite{stabilizer_inactivation} methods use 50-iteration serial BP2 in every stage.}
\label{fig:1270_and_882}
\end{figure*}

\section{Simulation}
\label{sec:simulation}
The message passing (MP) decoder used in this work is the sum-product (BP) version with flooding scheduling. All the computations are conducted using 32-bit floating-point precision. The LLR initialization for the first BP run uses a fixed $p=0.05$, independent of the real physical error rate. The messages from CN to VN in Eq (\ref{eq:X_VN_update},\ref{eq:Y_VN_update}) are multiplied by a normalization factor. After each GNN feedback, the BP takes this feedback as the channel LLR initialization, and starts with re-initialized messages. The \textbf{same} GNN weights are used in each round of feedback if there are multiple, see Alg. \ref{alg:sandwich_BP_GNN_eval} for the pseudo-code.

In \cite{stabilizer_inactivation}, the authors showed that SI outperformed OSD-0 when using \textbf{binary} message passing algorithms for both to obtain soft information for the post-processing. In the original work \cite{panteleev_degenerate}, the quaternary normalized min-sum (NMS) was used which shows a strong performance on the depolarizing channel (see Fig. (\ref{fig:1270_and_882}\textbf{b})). It is not clear how much improvement SI can gain when extended to the quaternary version, because its Tanner graph has unavoidable 4-cycles while the Tanner graph for $H_X$ does not, and SI depends on the convergence of MP on the remaining Tanner graph.

Recently in \cite{layered}, a randomly order layered scheduled MP decoder was proposed for the type of QLDPC codes we are investigating in this work, for example, the $\llbracket 882, 24\rrbracket$ code. This decoder involves no post-processing. CN updates are indexed into layers. The authors found that there exist 7 layers that could cover each check exactly twice for this code. When using a perturbed (multiply each CN message by a random factor in $\{0.875,0.9275\}$) normalized min-sum (NMS) as the MP decoder and randomizing the order of layer updates, they achieve very good results on the Z-noise channel (comparable to \textbf{binary} NMS+OSD in \cite{stabilizer_inactivation}). In addition, the decoding time is proportional to the $64\times 3.5=224$ steps for Z noise, where $64$ is the number of iterations, and the fractional layer number $3.5$ is because CNs in different layers cannot be updated in parallel.

In \cite{stabilizer_inactivation,layered}, the authors considered both sum-product (BP) and normalized min-sum (NMS) algorithms as message-passing decoders and various scheduling methods, we picked their best curves as a benchmark in Fig. (\ref{fig:1270_and_882}\textbf{b}). We are not competing with them as it is not fair for them, since they ignore the X and Z correlation when applying their algorithms. In fact, our methods could be used together, for example by replacing our flooding BP with their layered random order perturbed NMS decoder, though this will require retraining the feedback GNN.

\section{Conclusion}
\label{sec:conclusion}
We proposed an iterative decoding scheme for quantum LDPC codes based BP-4 and GNN decoding stages. We train the GNN layers such that they learn to refine the channel input LLRs for subsequent BP decoding stages. We demonstrated that our feedback GNN significantly lowers the error floor on two medium blocksize QLDPC codes when carefully trained. The feedback GNN benefits from a computational latency of $\mathcal{O}(1)$ when nodes are updated in parallel (and assuming a fixed number of iterations). However, in practice the complexity overhead of the feedback GNN is similar to a 16-iteration BP due to the computational complex matrix multiplications in the MLPs.
Furthermore, we proposed a new loss based on the boxplus operation that calculates a more accurate reliability estimate of the check nodes.
We leave it open for future research to improve the training strategy for faster convergence. Further, optimizing the decoder architecture for real-time implementations remains an open challenge for future work.


\begin{appendices}

\section{Loss Function}\label{sec:sine_vs_boxplus_loss}
Here we remark on the sine loss used in Neural BP \cite{NBP} and the boxplus loss used herein. 
As mentioned in the main text, the sine loss has some unwanted oscillation behavior. However, it can be used to train the Neural BP decoder by initializing such that it equals the classical BP decoder by setting all the weights to one. Hence the initialization of Neural BP is already a good decoder and the loss is already small. Reasonable training will not result in oscillating points due to the sine loss. However, when using this loss to train a full-GNN message-passing decoder from scratch (as done in \cite{cammerer2022graph}), we empirically observed that the loss does not converge. Therefore, we propose the boxplus loss as a more stable substitute.
Note that, as the channel posterior probabilities approach either $0$ or $1$, the sine loss is a good approximation to the boxplus loss, which is the analytical solution and involves the product of those posterior, whereas sine loss only involves the summation of those terms.

\begin{algorithm}
\SetKwInOut{KwIn}{Input}
\SetKwInOut{KwOut}{Output}
\SetAlgoLined\DontPrintSemicolon
\SetKwComment{Comment}{/* }{ */}
\SetKwFunction{sandwich}{Sandwich-BP-GNN-Decoder}\SetKwFunction{fbGNN}{Feedback-GNN}
\caption{Sandwiched BP GNN Decoding}\label{alg:sandwich_BP_GNN_eval} 
\SetKwProg{algo}{Algorithm}{}{}
\algo{\sandwich{decoders}}{
\KwIn{CSS code specified by $(H_X, H_Z)$ and a syndrome pair $(\bs_X, \bs_Z)$}
\KwOut{$\hbe=(\hbe_X,\hbe_Z)$}
\nl $\mathbf{\Lambda_0}\gets \log\frac{1-p}{p/3}\mathbf{I}_{N\times 3}$\;
\nl $(\hbe_X,\hbe_Z,\mathbf{\Lambda}_{\text{post}})\gets \text{decoders[0]}(\mathbf{\Lambda_0})$\;
\nl \If{$H_X\hbe_Z=\bs_Z$ and $H_Z\hbe_X=\bs_X$} {
\nl \KwRet $(\hbe_X,\hbe_Z)$\;
}
\Comment*[r]{Begin post-processing.}
\nl \ForEach{decoder in decoders[1:]} {
\nl $\mathbf{\Lambda'}\gets$ \fbGNN($\mathbf{\Lambda}_{\text{post}}$)\;
\nl $(\hbe_X,\hbe_Z,\mathbf{\Lambda}_{\text{post}})\gets \text{decoder}(\mathbf{\Lambda'})$\;
\nl \If{$H_X\hbe_Z=\bs_Z$ and $H_Z\hbe_X=\bs_X$} {
\nl \KwRet $(\hbe_X,\hbe_Z)$\;
}}
\nl \KwRet $(\hbe_X,\hbe_Z)$\;}{}
\end{algorithm}

\begin{algorithm}
\SetKwInOut{KwIn}{Input}
\SetKwInOut{KwOut}{Output}
\SetAlgoLined\DontPrintSemicolon
\SetKwComment{Comment}{/* }{ */}
\SetKwProg{proc}{Procedure}{}{}
\proc{\fbGNN{$\mathbf{\Lambda}_{\text{post}}$}}{
\nl \ForEach{VN $v_i$} {
\nl $\mathbf{h}_{v_i}\gets \mathbf{\Lambda}_{\text{post}, i}$\;
}
\nl \ForEach{CN $c_j$} {
\nl $h_{c_j}\gets\dots$\Comment*[r]{Eq (\ref{eq:combination_explicit},\ref{eq:CN_feature})}
}
\nl \ForEach{edge $(c_j,v_i)$} {
\nl $\mathbf{m}_{c_j\to v_i}\gets f^{C_{X/Z}\to V}(h_{c_j}||\mathbf{h}_{v_i})$\;
}
\nl \ForEach{VN $v_i$} {
\nl $\mathbf{m}_{v_i}^X\gets$ average of incoming messages from X-type CNs.\;
\nl $\mathbf{m}_{v_i}^Z\gets$ average of incoming messages from Z-type CNs.\;
\nl $\mathbf{\Lambda'}_i\gets f^{VN}(\mathbf{h}_{v_i}||\mathbf{m}_{v_i}^X||\mathbf{m}_{v_i}^Z)$\;
}
\nl \KwRet $\mathbf{\Lambda'}$\;}
\end{algorithm}
\noindent
\end{appendices}

{\small 
\setstretch{0.9}
\printbibliography[heading=bibintoc,title={\large References}]
}
\end{document}

%% file: macrosetup.tex
\newcommand{\R}{\mathbb{R}}

\newcommand{\bx}{\mathbf{x}}

\newcommand{\bs}{\mathbf{s}}
\newcommand{\be}{\mathbf{e}}

\newcommand{\hbe}{\hat{\mathbf{e}}}

\newcommand{\LX}{\Lambda^{X}}
\newcommand{\LY}{\Lambda^{Y}}
\newcommand{\LZ}{\Lambda^{Z}}

\newcommand{\wt}{\text{wt}}
\newcommand{\im}{\text{im\ }}

\newcommand{\boxplusop}{
  \mathop{
    \vphantom{\oplus} 
    \mathchoice
      {\vcenter{\hbox{\resizebox{\widthof{$\displaystyle\oplus$}}{!}{$\boxplus$}}}}
      {\vcenter{\hbox{\resizebox{\widthof{$\bigoplus$}}{!}{$\boxplus$}}}}
      {\vcenter{\hbox{\resizebox{\widthof{$\scriptstyle\oplus$}}{!}{$\boxplus$}}}}
      {\vcenter{\hbox{\resizebox{\widthof{$\scriptscriptstyle\oplus$}}{!}{$\boxplus$}}}}
  }\displaylimits 
}

\renewcommand{\epsilon}{\ensuremath\varepsilon}

\renewcommand{\phi}{\ensuremath{\varphi}}

%% file: block_diagram.tex
\begin{tikzpicture}
\node [draw,
    fill=lemon,
    minimum width=1.0cm,
    minimum height=1.2cm,
]  (BP1) {64};
\node[below] at (BP1.south){\tiny BP};

\coordinate[left of=BP1] (d1);

\node [draw,
    fill=coral,
    minimum width=0.1cm,
    minimum height=1.2cm,
    right=0.6cm of BP1,
]  (GNN1) {1};
\node[below] at (GNN1.south){\tiny GNN};

\node [draw,
    fill=lemon,
    minimum width=0.1cm,
    minimum height=1.2cm,
    right=0.4cm of GNN1,
]  (BP2) {16};
\node[below] at (BP2.south){\tiny BP};

\node [draw,
    fill=coral,
    minimum width=0.1cm,
    minimum height=1.2cm,
    right=0.6cm of BP2,
]  (GNN2) {1};
\node[below] at (GNN2.south){\tiny GNN};

\node [draw,
    fill=lemon,
    minimum width=0.1cm,
    minimum height=1.2cm,
    right=0.4cm of GNN2,
]  (BP3) {16};
\node[below] at (BP3.south){\tiny BP};

\node [draw,
    fill=coral,
    minimum width=0.1cm,
    minimum height=1.2cm,
    right=0.6cm of BP3,
]  (GNN3) {1};
\node[below] at (GNN3.south){\tiny GNN};

\node [draw,
    fill=lemon,
    minimum width=0.1cm,
    minimum height=1.2cm,
    right=0.4cm of GNN3,
]  (BP4) {16};
\node[below] at (BP4.south){\tiny BP};
\coordinate[right=0.6cm of BP4] (d2);

\draw[-stealth] (d1) -- (BP1.west)
    node[midway,above=0.03cm]{\tiny$\mathbf{\Lambda}_0$};
\draw[-stealth] (BP1.east) -- (GNN1.west)
    node[midway,above]{\tiny$\mathbf{\Lambda}_{\text{post}}$};
\draw[-stealth] (GNN1.east) -- (BP2.west)
    node[midway,above=0.06cm]{\tiny$\mathbf{\Lambda'}$};
\draw[-stealth] (BP2.east) -- (GNN2.west)
    node[midway,above]{\tiny$\mathbf{\Lambda}_{\text{post}}$};
\draw[-stealth] (GNN2.east) -- (BP3.west)
    node[midway,above=0.06cm]{\tiny$\mathbf{\Lambda'}$};
\draw[-stealth] (BP3.east) -- (GNN3.west)
    node[midway,above]{\tiny$\mathbf{\Lambda}_{\text{post}}$};
\draw[-stealth] (GNN3.east) -- (BP4.west)
    node[midway,above=0.06cm]{\tiny$\mathbf{\Lambda'}$};
\draw[-stealth] (BP4.east) -- (d2)
    node[midway,above]{\tiny$\mathbf{\Lambda}_{\text{post}}$};
\end{tikzpicture}

%% file: fb_gnn.tex
\definecolor{mittelblau}{RGB}{0, 126, 198}
\definecolor{violettblau}{cmyk}{0.9, 0.6, 0, 0}
\definecolor{rot}{RGB}{238, 28 35}
\definecolor{apfelgruen}{RGB}{140, 198, 62}
\definecolor{gelb}{RGB}{1, 221, 0}
\definecolor{orange}{RGB}{244, 111, 33}
\definecolor{pink}{RGB}{237, 0, 140}
\definecolor{lila}{RGB}{128, 10, 145}
\definecolor{hellgrau}{RGB}{224, 224, 224}
\definecolor{mittelgrau}{RGB}{128, 128, 128}
\definecolor{dunkelgrau}{RGB}{80,80,80}
\definecolor{anthrazit}{RGB}{19, 31, 31}
\definecolor{lavender}{RGB}{227,159,246}
\definecolor{mintgreen}{RGB}{170,240,209}
\definecolor{lemonade}{RGB}{252,186,203}
\definecolor{lemon}{RGB}{255,247,0}
\definecolor{darkcyan}{RGB}{0,139,139}

\usetikzlibrary{shapes.geometric}
\tikzstyle{cndx} = [rectangle,draw=rot,fill=white,thick,text width=2mm,text height=2mm,inner sep=0pt]
\tikzstyle{cndz} = [rectangle,draw=mittelblau,fill=white,thick,text width=2mm,text height=2mm,inner sep=0pt]

\tikzstyle{vnd} = [circle,draw=black,fill=white,thick,text width=2mm,inner sep=0pt]

\tikzstyle{lam} = [ellipse, minimum width=1cm, minimum height=0.5cm,,draw=black,fill=white,text width=2mm,inner sep=0pt]



\tikzstyle{lambda} = [rectangle,minimum width=0.5cm, minimum height=4.8cm,draw=teal,fill=none,thick,text width=0.5cm,text height=4.5mm,inner sep=0pt]

\tikzstyle{lambdaout} = [rectangle,minimum width=0.5cm, minimum height=4.8cm,draw=lavender,fill=none,thick,text width=0.5cm,text height=4.5mm,inner sep=0pt]

\tikzstyle{outerbox} = [rectangle,minimum width=5.6cm, minimum height=5.4cm,draw=coral,fill=none,thick,text width=0.5cm,text height=4.5mm,inner sep=0pt]

\tikzstyle{fvn} = [rectangle,draw=black,fill=black,thick,text width=1mm,text height=1mm,inner sep=0pt]

\tikzstyle{fc2vx} = [rectangle,draw=rot,fill=rot,thick,text width=1mm,text height=1mm,inner sep=0pt]

\tikzstyle{fc2vz} = [rectangle,draw=mittelblau,fill=mittelblau,thick,text width=1mm,text height=1mm,inner sep=0pt]

\begin{tikzpicture}[scale=0.6]

	    \node [vnd, label={[yshift=-0.02cm,xshift=0.cm]$v_1$}] (vnd1) {};
	    \node [vnd, below=0.5cm of vnd1, label={[yshift=-0.02cm,xshift=-0.cm]$v_2$}] (vnd2) {};
	    \node [vnd, below=0.5cm of vnd2, label={[yshift=-0.02cm,xshift=-0.cm]$v_3$}] (vnd3) {};
	    \node [vnd, below=0.5cm of vnd3, label={[yshift=-0.02cm,xshift=-0.cm]$v_4$}] (vnd4) {};
	    \node [vnd, below=0.5cm of vnd4, label={[yshift=-0.02cm,xshift=-0.cm]$v_5$}] (vnd5) {};
	    \node [vnd, below=0.5cm of vnd5, label={[yshift=-0.02cm,xshift=-0.cm]$v_6$}] (vnd6) {};
	    \node [vnd, below=0.5cm of vnd6, label={[yshift=-0.02cm,xshift=-0.cm]$v_7$}] (vnd7) {};

        \node [lambda, above left=-2.5cm and 5cm of vnd4, label={[yshift=-2.6cm, xshift=0.cm]{\tiny$\mathbf{\Lambda}_{\text{post}}$}}](lambda_in) {};
        
        \node [lambdaout, above right=-2.5cm and 1cm of vnd4, label={[yshift=-2.6cm, xshift=0.cm]{\tiny$\mathbf{\Lambda'}$}}](lambda_out) {};
              
	    \node [cndx, below left=-0.1cm and 1.5cm of vnd1, label={[yshift=0.0cm,xshift=0.cm]$c_1$}] (cnd1) {};
	    \node [cndx, below=0.6cm of cnd1, label={[yshift=0.0cm,xshift=0.cm]$c_2$}] (cnd2) {};
	    \node [cndx, below=0.6cm of cnd2, label={[yshift=0.0cm,xshift=0.cm]$c_3$}] (cnd3) {};
        \node [cndz, below=0.6cm of cnd3, label={[yshift=0.0cm,xshift=0.cm]$c_4$}] (cnd4) {};
	    \node [cndz,below=0.6cm of cnd4, label={[yshift=0.0cm,xshift=0.cm]$c_5$}] (cnd5) {};
        \node [cndz, below=0.6cm of cnd5, label={[yshift=0.0cm,xshift=0.cm]$c_6$}] (cnd6) {};

	    \node [lam, left=3cm of vnd1, label={[yshift=-0.55cm,xshift=0.cm]\tiny$\lambda_1^{X,Z}$}] (lam1) {};
	    \node [lam, left=3cm of vnd2, label={[yshift=-0.55cm,xshift=-0.cm]\tiny$\lambda_2^{X,Z}$}] (lam2) {};
	    \node [lam, left=3cm of vnd3, label={[yshift=-0.55cm,xshift=-0.cm]\tiny$\lambda_3^{X,Z}$}] (lam3) {};
	    \node [lam, left=3cm of vnd4, label={[yshift=-0.55cm,xshift=-0.cm]\tiny$\lambda_4^{X,Z}$}] (lam4) {};
	    \node [lam, left=3cm of vnd5, label={[yshift=-0.55cm,xshift=-0.cm]\tiny$\lambda_5^{X,Z}$}] (lam5) {};
	    \node [lam, left=3cm of vnd6, label={[yshift=-0.55cm,xshift=-0.cm]\tiny$\lambda_6^{X,Z}$}] (lam6) {};
	    \node [lam, left=3cm of vnd7, label={[yshift=-0.55cm,xshift=-0.cm]\tiny$\lambda_7^{X,Z}$}] (lam7) {};


        \path[-,red] (vnd1) edge node [fc2vx, label={[xshift=-0.cm,yshift=-0.1cm]\tiny \textcolor{rot}{$f^{C_X\to V}$}}] {} (cnd1);
        \path[-,red] (vnd3) edge node [fc2vx] {} (cnd1);
        \path[-,red] (vnd5) edge node [fc2vx] {} (cnd1);
        \path[-,red] (vnd7) edge node [fc2vx] {} (cnd1);
        
        \path[-,red] (lam1) edge node [yshift=0.2cm] {\tiny \textcolor{black}{Eq (\ref{eq:CN_feature})}} (cnd1);
        \path[-,red] (lam3.east) edge node{} (cnd1);
        \path[-,red] (lam5.east) edge node{} (cnd1);
        \path[-,red] (lam7.east) edge node{} (cnd1);

	  \path[-,red] (vnd2) edge node [fc2vx] {} (cnd2);
	  \path[-,red] (vnd3) edge node [fc2vx] {} (cnd2);
	  \path[-,red] (vnd6) edge node [fc2vx] {} (cnd2);
	  \path[-,red] (vnd7) edge node [fc2vx] {} (cnd2);
   
   	\path[-,red] (lam2.east) edge node{} (cnd2);
	  \path[-,red] (lam3.east) edge node{} (cnd2);
	  \path[-,red] (lam6.east) edge node{} (cnd2);
	  \path[-,red] (lam7.east) edge node{} (cnd2);

	  \path[-,red] (vnd4) edge node [fc2vx] {} (cnd3);
	  \path[-,red] (vnd5) edge node [fc2vx] {} (cnd3);
	  \path[-,red] (vnd6) edge node [fc2vx] {} (cnd3);
	  \path[-,red] (vnd7) edge node [fc2vx] {} (cnd3);

	  \path[-,red] (lam4.east) edge node{} (cnd3);
	  \path[-,red] (lam5.east) edge node{} (cnd3);
	  \path[-,red] (lam6.east) edge node{} (cnd3);
	  \path[-,red] (lam7.east) edge node{} (cnd3);

	  \path[-,mittelblau] (vnd1) edge node [fc2vz] {} (cnd4);
	  \path[-,mittelblau] (vnd3) edge node [fc2vz] {} (cnd4);
	  \path[-,mittelblau] (vnd5) edge node [fc2vz] {} (cnd4);
	  \path[-,mittelblau] (vnd7) edge node [fc2vz] {} (cnd4);

	  \path[-,mittelblau] (lam1.east) edge node{} (cnd4);
	  \path[-,mittelblau] (lam3.east) edge node{} (cnd4);
	  \path[-,mittelblau] (lam5.east) edge node{} (cnd4);
	  \path[-,mittelblau] (lam7.east) edge node{} (cnd4);

	  \path[-,mittelblau] (vnd2) edge node [fc2vz] {} (cnd5);
	  \path[-,mittelblau] (vnd3) edge node [fc2vz] {} (cnd5);
	  \path[-,mittelblau] (vnd6) edge node [fc2vz] {} (cnd5);
	  \path[-,mittelblau] (vnd7) edge node [fc2vz] {} (cnd5);

	  \path[-,mittelblau] (lam2.east) edge node {} (cnd5);
	  \path[-,mittelblau] (lam3.east) edge node {} (cnd5);
	  \path[-,mittelblau] (lam6.east) edge node {} (cnd5);
	  \path[-,mittelblau] (lam7.east) edge node {} (cnd5);

	  \path[-,mittelblau] (vnd4) edge node [fc2vz] {} (cnd6);
	  \path[-,mittelblau] (vnd5) edge node [fc2vz] {} (cnd6);
	  \path[-,mittelblau] (vnd6) edge node [fc2vz] {} (cnd6);
	  \path[-,mittelblau] (vnd7) edge node [fc2vz,label={[xshift=-0.cm,yshift=-0.5cm]\tiny \textcolor{mittelblau}{$f^{C_Z\to V}$}}] {} (cnd6);

	  \path[-,mittelblau] (lam4.east) edge node {} (cnd6);
	  \path[-,mittelblau] (lam5.east) edge node {} (cnd6);
	  \path[-,mittelblau] (lam6.east) edge node {} (cnd6);
	  \path[-,mittelblau] (lam7.east) edge node {} (cnd6);

        \path[-stealth,black] ([yshift=3.66cm]lambda_in.east) edge node [xshift=0.2cm,yshift=0.2cm] {\tiny Eq (\ref{eq:combination_explicit})} (lam1.west);
        \path[-stealth,black] ([yshift=2.44cm]lambda_in.east) edge (lam2.west);
        \path[-stealth,black] ([yshift=1.22cm]lambda_in.east) edge (lam3.west);
        \path[-stealth,black] ([yshift=0cm]lambda_in.east) edge (lam4.west);
        \path[-stealth,black] ([yshift=-1.22cm]lambda_in.east) edge (lam5.west);
        \path[-stealth,black] ([yshift=-2.44cm]lambda_in.east) edge (lam6.west);
        \path[-stealth,black] ([yshift=-3.66cm]lambda_in.east) edge (lam7.west);

        \path[-stealth,black] (vnd1) edge node [fvn, label={[xshift=-0.1cm,yshift=-0.1cm]\tiny $f^{VN}$}] {} ([yshift=3.66cm]lambda_out.west);
        \path[-stealth,black] (vnd2) edge node [fvn] {} ([yshift=2.44cm]lambda_out.west);
        \path[-stealth,black] (vnd3) edge node [fvn] {} ([yshift=1.22cm]lambda_out.west);
        \path[-stealth,black] (vnd4) edge node [fvn] {} ([yshift=0cm]lambda_out.west);
        \path[-stealth,black] (vnd5) edge node [fvn] {} ([yshift=-1.22cm]lambda_out.west);
        \path[-stealth,black] (vnd6) edge node [fvn] {} ([yshift=-2.44cm]lambda_out.west);
        \path[-stealth,black] (vnd7) edge node [fvn] {} ([yshift=-3.66cm]lambda_out.west);

        \node [outerbox, above left=-2.65cm and -0.9cm of vnd4, label={[yshift=0.05cm,xshift=0.cm]\textcolor{coral}{Unrolled feedback GNN}}] (outer_box) {};

\end{tikzpicture}

%% file: n1270_progression.tex
\begin{tikzpicture}
{\small
\definecolor{blueviolet}{RGB}{138,43,226}
\definecolor{darkgray176}{RGB}{176,176,176}
\definecolor{lightgray211}{RGB}{211,211,211}
\definecolor{darkorange25512714}{RGB}{255,127,14}
\definecolor{dodgerblue}{RGB}{30,144,255}
\definecolor{gold}{RGB}{255,215,0}
\definecolor{red}{RGB}{255,0,0}
\definecolor{orange}{RGB}{255,93,0}
\definecolor{limegreen}{RGB}{50,205,50}
\definecolor{steelblue31119180}{RGB}{31,119,180}
\definecolor{teal}{RGB}{0,128,128}
\definecolor{olivegreen}{RGB}{55,126,34}
\definecolor{neonpurple}{RGB}{176,38,255}
\definecolor{coral}{RGB}{254,125,106}
\definecolor{lemonade}{RGB}{252,186,203}
\definecolor{lemon}{RGB}{255,247,0}
\definecolor{amber}{RGB}{255,191,0}
\definecolor{palegreen}{RGB}{180,238,180}
\definecolor{lightteal}{RGB}{64,224,208}
\definecolor{lavender}{RGB}{227,159,246}
\definecolor{darkviolet}{RGB}{153,50,204}
\definecolor{darkcyan}{RGB}{0,139,139}
\definecolor{darkgreen}{RGB}{0,100,0}
\definecolor{lightskyblue}{RGB}{136,205,250}

\begin{axis}[
width=0.95\textwidth,
log basis y={10},
tick align=inside,
tick pos=left,
x grid style={lightgray211},
xlabel={Physical error rate \(\displaystyle p\)},
xmajorgrids,
xmode=log,
xmin=0.02, xmax=0.14,
xtick={0.02,0.1,0.14},
xticklabels={
  \(\displaystyle {0.02}\),
  \(\displaystyle {0.1}\),
  \(\displaystyle {0.14}\),
},
minor xtick={0.03,0.04,0.05,0.06,0.07,0.08,0.09},
xminorgrids=true,
scaled x ticks=false,
xtick style={color=black},
y grid style={lightgray211},
ylabel={Logical error rate \(\displaystyle P_L\)},
ymajorgrids,
ymin=1e-8, ymax=1,
ymode=log,
semithick,
ytick style={color=black},
ytick={1e-8,1e-7,1e-6,1e-5,0.0001,0.001,0.01,0.1,1,10,100},
yticklabels={
  \(\displaystyle {10^{-8}}\),
  \(\displaystyle {10^{-7}}\),
  \(\displaystyle {10^{-6}}\),
  \(\displaystyle {10^{-5}}\),
  \(\displaystyle {10^{-4}}\),
  \(\displaystyle {10^{-3}}\),
  \(\displaystyle {0.01}\),
  \(\displaystyle {0.1}\),
  \(\displaystyle {1}\),
},
error bars/y dir=both,
error bars/y explicit,
legend columns=6,
transpose legend,
legend style={at={(0,1)}, anchor=north west, font=\tiny, fill opacity=0.8, draw opacity=0.8, text opacity=1, draw=none, /tikz/every even column/.append style={column sep=-0.5cm}},
legend cell align=left,
]

\addlegendentry{\textbf{BP iterations}}
\addlegendimage{empty legend}

\addplot [amber, dotted, mark=diamond*, mark size=2.0, mark options={solid}, forget plot]
table [x=P, y=WER] {feedback_GNN_coarse_1G_factor08.dat};
\addplot [amber, mark=diamond*, mark size=2.0, mark options={solid}]
table [x=P, y=WER] {feedback_GNN_coarse_1G.dat};
\addlegendentry{coarse $(64,16)$}

\addplot [palegreen, dotted, mark=diamond*, mark size=2.0, mark options={solid}, forget plot]
table [x=P, y=WER] {feedback_GNN_coarse_1G_64_64_factor08.dat};
\addplot [palegreen, mark=diamond*, mark size=2.0, mark options={solid}]
table [x=P, y=WER] {feedback_GNN_coarse_1G_64_64.dat};
\addlegendentry{coarse $(64,64)$}

\addplot [lavender, dotted, mark=diamond*, mark size=2.0, mark options={solid}, forget plot]
table [x=P, y=WER] {feedback_GNN_coarse_3G_factor08.dat};
\addplot [lavender, mark=diamond*, mark size=2.0, mark options={solid}]
table [x=P, y=WER] {feedback_GNN_coarse_3G.dat};
\addlegendentry{coarse $(64,16,16,16)$}

\addplot [lightskyblue, dotted, mark=*, mark size=0.5, mark options={solid}, forget plot]
table [x=P, y=WER] {flooding_BP_64_factor08.dat};
\addplot [lightskyblue, mark=*, mark size=0.5, mark options={solid}]
table [x=P, y=WER] {flooding_BP_64.dat};
\addlegendentry{BP only $64$}

\addplot [darkcyan, dotted, mark=*, mark size=0.5, mark options={solid}, forget plot]
table [x=P, y=WER] {flooding_BP_128_factor08.dat};
\addplot [darkcyan, mark=*, mark size=0.5, mark options={solid}]
table [x=P, y=WER] {flooding_BP_128.dat};
\addlegendentry{BP only $128$}

\addplot [coral, dotted, mark=*, mark size=0.5, mark options={solid}, forget plot]
table [x=P, y=WER] {feedback_GNN_1G_factor08.dat};
\addplot [coral, mark=*, mark size=0.5, mark options={solid}]
table [x=P, y=WER] {feedback_GNN_1G.dat};
\addlegendentry{refined $(64,16)$}

\addplot [lightteal, dotted, mark=*, mark size=0.5, mark options={solid}, forget plot]
table [x=P, y=WER] {feedback_GNN_1G_64_64_factor08.dat};
\addplot [lightteal, mark=*, mark size=0.5, mark options={solid}]
table [x=P, y=WER] {feedback_GNN_1G_64_64.dat};
\addlegendentry{refined $(64,64)$}

\addplot [darkviolet, dotted, mark=*, mark size=0.5, mark options={solid}, forget plot]
table [x=P, y=WER] {feedback_GNN_3G_factor08_repeat.dat};
\addplot [darkviolet, mark=*, mark size=0.5, mark options={solid}]
table [x=P, y=WER] {feedback_GNN_3G_repeat.dat};
\addlegendentry{refined $(64,16,16,16)$}

\addlegendentry{\textbf{Method}}
\addlegendimage{empty legend}
\addlegendentry{$\kappa=1.0$}
\addlegendimage{black, line legend}
\addlegendentry{$\kappa=0.8$}
\addlegendimage{black, line legend, dotted}

\end{axis}
}
\end{tikzpicture}

%% file: n1270_k28.tex
\begin{tikzpicture}
{\small
\definecolor{blueviolet}{RGB}{138,43,226}
\definecolor{darkgray176}{RGB}{176,176,176}
\definecolor{lightgray211}{RGB}{211,211,211}
\definecolor{darkorange25512714}{RGB}{255,127,14}
\definecolor{dodgerblue}{RGB}{30,144,255}
\definecolor{gold}{RGB}{255,215,0}
\definecolor{red}{RGB}{255,0,0}
\definecolor{orange}{RGB}{255,93,0}
\definecolor{limegreen}{RGB}{50,205,50}
\definecolor{steelblue31119180}{RGB}{31,119,180}
\definecolor{teal}{RGB}{0,128,128}
\definecolor{olivegreen}{RGB}{55,126,34}
\definecolor{darkviolet}{RGB}{153,50,204}


\begin{axis}[
width=0.475\textwidth,
log basis y={10},
tick align=inside,
tick pos=left,
x grid style={lightgray211},
xlabel={Physical error rate \(\displaystyle p\)},
xmajorgrids,
xmode=log,
xmin=0.02, xmax=0.14,
xtick={0.02,0.1,0.14},
xticklabels={
  \(\displaystyle {0.02}\),
  \(\displaystyle {0.1}\),
  \(\displaystyle {0.14}\),
},
minor xtick={0.03,0.04,0.05,0.06,0.07,0.08,0.09},
xminorgrids=true,
scaled x ticks=false,
xtick style={color=black},
y grid style={lightgray211},
ylabel={Logical error rate \(\displaystyle P_L\)},
ymajorgrids,
ymin=1e-8, ymax=1,
ymode=log,
semithick,
ytick style={color=black},
ytick={1e-8,1e-7,1e-6,1e-5,0.0001,0.001,0.01,0.1,1,10,100},
yticklabels={
  \(\displaystyle {10^{-8}}\),
  \(\displaystyle {10^{-7}}\),
  \(\displaystyle {10^{-6}}\),
  \(\displaystyle {10^{-5}}\),
  \(\displaystyle {10^{-4}}\),
  \(\displaystyle {10^{-3}}\),
  \(\displaystyle {0.01}\),
  \(\displaystyle {0.1}\),
  \(\displaystyle {1}\),
},
error bars/y dir=both,
error bars/y explicit,
legend columns=2,
legend style={at={(0,1)}, anchor=north west, font=\tiny, fill opacity=0.8, draw opacity=0.8, text opacity=1, draw=none},
legend cell align=left,
]%
\addlegendentry{\textbf{Method, $N_a$}}
\addlegendimage{empty legend}

\addplot [black, mark=*, mark size=0.5, mark options={solid}]
table [x=P, y=WER] {BP_OSD0.dat};
\addlegendentry{\cite{panteleev_degenerate} NMS+OSD0}
\addplot [cyan, mark=*, mark size=0.5, mark options={solid}]
table [x=P, y=WER] {BP_pert_retry100.dat};
\addlegendentry{\cite{perturbation} Pert, $100$}
\addplot [black, dashed, mark=*, mark size=0.5, mark options={solid}]
table [x=P, y=WER] {layered_BP.dat};
\addlegendentry{\cite{panteleev_degenerate} Layered NMS}
\addplot [olivegreen, mark=*, mark size=0.5, mark options={solid}]
table [x=P, y=WER] {BP_aug_retry100.dat};
\addlegendentry{\cite{augmentation} Aug, $100$}
\addplot [olivegreen, dashed, mark=*, mark size=0.5, mark options={solid}]
table [x=P, y=WER] {BP_aug_retry10.dat};
\addlegendentry{\cite{augmentation} Aug, $10$}
\addplot [blue, mark=*, mark size=0.5, mark options={solid}]
table [x=P, y=WER] {BP_enh_retry100.dat};
\addlegendentry{\cite{enhanced_feedback} Enh, $100$}
\addplot [blue, dashed, mark=*, mark size=0.5, mark options={solid}]
table [x=P, y=WER] {BP_enh_retry10.dat};
\addlegendentry{\cite{enhanced_feedback} Enh, $10$}

\addplot [darkviolet, dotted, mark=*, mark size=0.5, mark options={solid}]
table [x=P, y=WER] {feedback_GNN_3G_factor08.dat};
\addlegendentry{GNN $0.8$, $3$}
\addplot [darkviolet, mark=*, mark size=0.5, mark options={solid}]
table [x=P, y=WER] {feedback_GNN_3G.dat};
\addlegendentry{GNN $1.0$, $3$}
\addplot [red, dotted, mark=*, mark size=0.5, mark options={solid}]
table [x=P, y=WER] {feedback_GNN_5G_factor08.dat};
\addlegendentry{GNN $0.8$, $5$}
\addplot [red, mark=*, mark size=0.5, mark options={solid}]
table [x=P, y=WER] {feedback_GNN_5G.dat};
\addlegendentry{GNN $1.0$, $5$}

\end{axis}
};%
\node [below right=1cm and 1cm]
{(\textbf{a}) $\llbracket 1270,28, \leq 46\rrbracket$};
\end{tikzpicture}%

%% file: n882_k24.tex
\begin{tikzpicture}
{\small
\definecolor{blueviolet}{RGB}{138,43,226}
\definecolor{darkgray176}{RGB}{176,176,176}
\definecolor{lightgray211}{RGB}{211,211,211}
\definecolor{darkorange25512714}{RGB}{255,127,14}
\definecolor{dodgerblue}{RGB}{30,144,255}
\definecolor{gold}{RGB}{255,215,0}
\definecolor{red}{RGB}{255,0,0}
\definecolor{orange}{RGB}{255,93,0}
\definecolor{limegreen}{RGB}{50,205,50}
\definecolor{steelblue31119180}{RGB}{31,119,180}
\definecolor{teal}{RGB}{0,128,128}
\definecolor{olivegreen}{RGB}{55,126,34}
\definecolor{darkviolet}{RGB}{153,50,204}

\begin{axis}[
width=0.475\textwidth,
log basis y={10},
tick align=inside,
tick pos=left,
x grid style={lightgray211},
xlabel={Physical error rate \(\displaystyle p\)},
xmajorgrids,
xmode=log,
xmin=0.04, xmax=0.14,
xtick={0.04,0.1,0.14},
xticklabels={
  \(\displaystyle {0.04}\),
  \(\displaystyle {0.1}\),
  \(\displaystyle {0.14}\),
},
minor xtick={0.05,0.06,0.07,0.08,0.09},
xminorgrids=true,
scaled x ticks=false,
xtick style={color=black},
y grid style={lightgray211},
ylabel={Logical error rate \(\displaystyle P_L\)},
ymajorgrids,
ymin=1e-7, ymax=1,
ymode=log,
semithick,
ytick style={color=black},
ytick={1e-7,1e-6,1e-5,0.0001,0.001,0.01,0.1,1,10,100},
yticklabels={
  \(\displaystyle {10^{-7}}\),
  \(\displaystyle {10^{-6}}\),
  \(\displaystyle {10^{-5}}\),
  \(\displaystyle {10^{-4}}\),
  \(\displaystyle {10^{-3}}\),
  \(\displaystyle {0.01}\),
  \(\displaystyle {0.1}\),
  \(\displaystyle {1}\),
},
error bars/y dir=both,
error bars/y explicit,
legend columns=2,
legend style={at={(0,1)}, anchor=north west, font=\tiny, fill opacity=0.8, draw opacity=0.8, text opacity=1, draw=none},
legend cell align=left,
]%
\addlegendentry{\textbf{Method, $N_a$}}
\addlegendimage{empty legend}

\addplot [black, mark=*, mark size=0.5, mark options={solid}]
table [x=P, y=WER] {BP_OSD0_n882.dat};
\addlegendentry{\cite{panteleev_degenerate} NMS+OSD0}
\addplot [black, dashed, mark=*, mark size=0.5, mark options={solid}]
table [x=P, y=WER] {layered_BP_n882.dat};
\addlegendentry{\cite{panteleev_degenerate} Layered NMS}
\addplot [cyan, mark=*, mark size=0.5, mark options={solid}]
table [x=P, y=WER] {layered_ro_NMS.dat};
\addlegendentry{\cite{layered} Layered ro NMS}
\addplot [olivegreen, mark=*, mark size=0.5, mark options={solid}]
table [x=P, y=WER] {S_SP_SI_10.dat};
\addlegendentry{\cite{stabilizer_inactivation} SI, 10}
\addplot [olivegreen, dashed, mark=*, mark size=0.5, mark options={solid}]
table [x=P, y=WER] {S_SP_SI_all.dat};
\addlegendentry{\cite{stabilizer_inactivation} SI, $\geq 429$}
\addplot [darkviolet, dotted, mark=*, mark size=0.5, mark options={solid}]
table [x=P, y=WER] {feedback_GNN_3G_factor08_882.dat};
\addlegendentry{GNN $0.8$, $3$}
\addplot [darkviolet, mark=*, mark size=0.5, mark options={solid}]
table [x=P, y=WER] {feedback_GNN_3G_882.dat};
\addlegendentry{GNN $1.0$, $3$}
\addplot [red, dotted, mark=*, mark size=0.5, mark options={solid}]
table [x=P, y=WER] {feedback_GNN_5G_factor08_882.dat};
\addlegendentry{GNN $0.8$, $5$ }
\addplot [red, mark=*, mark size=0.5, mark options={solid}]
table [x=P, y=WER] {feedback_GNN_5G_882.dat};
\addlegendentry{GNN $1.0$, $5$ }

\end{axis}
};%
\node [below right=1cm and 1cm]
{(\textbf{b}) $\llbracket 882,24,\leq 24 \rrbracket$};
\end{tikzpicture}%